\documentclass[prl,twocolumn,showpacs,groupedaddress]{revtex4}
\usepackage{graphicx}
\usepackage{amsmath}
\graphicspath{{./img/}}
\newcommand{\sect}[2]{{\par\it #1.---}{#2}}
\begin{document}
\title{Current Noise in ac-Driven Nanoscale Conductors}
\author{S\'ebastien Camalet}
\author{J\"org Lehmann}
\author{Sigmund Kohler}
\author{Peter H\"anggi}
\affiliation{Institut f\"ur Physik, Universit\"at Augsburg,
        Universit\"atsstra\ss e~1, D-86135 Augsburg, Germany}
\date{\today}
%
\begin{abstract}

The theory for current fluctuations in ac-driven transport through
nanoscale systems is put forward. By use of a generalized,
non-Hermitian Floquet theory we derive novel explicit expressions for the
time-averaged current and the zero-frequency component
of the power spectrum of current fluctuations.  A distinct suppression
of both the zero-frequency noise and the dc-current occurs for
suitably tailored ac-fields. The relative level of transport noise,
being characterized by a Fano factor, can selectively be manipulated
by ac-sources; in particular, it exhibits both characteristic maxima
and minima near current suppression.

\end{abstract}
\maketitle


Recent experimental successes in the coherent coupling of quantum dots
\cite{Blick1996a} and in the reproducible measurement of electronic
currents through molecules \cite{Cui2001a, Reichert2002a} have given
rise to renewed theoretical interest in the transport properties of
nanoscale systems \cite{Nitzan2001a, Hanggi2002a}. Thereby, new
ideas in order to exploit the quantum coherence of
such systems for the construction of novel electronic devices
\cite{Hanggi2002a} have emerged. One possible construction element is
based on the manipulation of quantum dots or single molecules by
use of an oscillating gate voltage or an infrared laser, respectively.
A prominent effect of such ac-fields consists in the adiabatic
\cite{Thouless1983a, Brouwer1998a, Altshuler1999a, Switkes1999a} and
nonadiabatic \cite{Wagner1999a,Wang2002a} pumping of electrons.
Moreover, laser irradiated molecular wires provide novel devices such
as coherent quantum rectifiers \cite{Lehmann2002b} and optically
controlled transistors \cite{Lehmann2002c}.
However, such time-dependent control schemes can be valuable in
practice only if they operate at tolerable noise levels. Thus, the
question whether noise properties of nanoscale systems can be
selectively manipulated becomes of foremost interest.

Electron transport through time-independent, mesoscopic systems is
commonly described within the framework of a scattering formalism.
Both the average current \cite{Datta1995a} and the transport noise
characteristics \cite{Blanter2000a, Chen2002a} can be expressed in
terms of the quantum transmission coefficients for the corresponding
transport channels.
By contrast, the theory for driven quantum transport is much less
developed.  Expressions for the spectral density of the current
fluctuations have been derived for the low-frequency ac-conductance
\cite{Pedersen1998a} and the scattering by a slowly time-dependent
potential \cite{Lesovik1994a}.
However, the situation becomes more opaque in the presence of rapidly
varying time-dependent fields.
Within a Green function approach, a \textit{formal} expression for the
current through a time-dependent conductor has been presented in
Refs.~\cite{Datta1992a, Jauho1994a}.
Here, we derive \textit{explicit} expressions for both the current and the
noise properties of electron transport through a nanoscale
conductor under the influence of time-dependent forces at
arbitrary frequency and strength. The dynamics of the electrons
is solved by integrating the Heisenberg equations of motion for
the electron creation/annihilation operators within a generalized
Floquet approach. We then use the resulting expressions to explore
the possibility of an \textit{a priori} control of the dc-current
and the zero-frequency noise by the influence of an ac-field.

\begin{figure}[t]
\includegraphics[width=0.85\columnwidth]{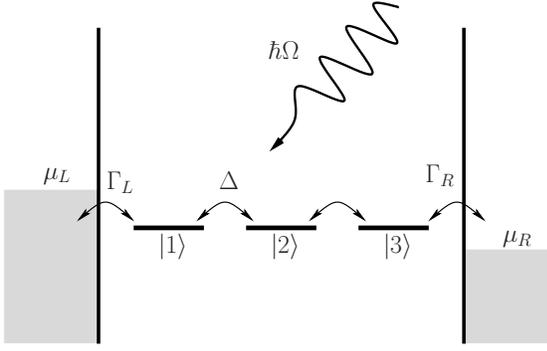}
\caption{\label{fig:levels} Level structure of the molecular wire with
$N=3$ orbitals. The end sites are coupled to two leads with chemical
potentials $\mu_L$ and $\mu_R=\mu_L-eV$.}
\end{figure}

\sect{The lead-wire model}
The entire setup of our nano\-scale system is described by the
time-dependent Hamiltonian $H(t) = H_\mathrm{wire}(t) + H_{\rm leads}
+ H_{\rm contacts}$, where the different terms correspond to the
driven wire (or coupled quantum dots), the leads, and the wire-leads
coupling, respectively.  In order to go beyond merely formal
considerations, we herewith focus on the regime of \textit{coherent
quantum transport} where the main physics at work occurs on the wire
itself.  In doing so, we neglect other possible influences stemming
from driving induced hot electrons in the leads, dissipation on the
wire and, as well, electron-electron interaction effects.  Then, the
wire Hamiltonian reads in a tight-binding approximation with
$N$ orbitals $|n\rangle$
\begin{equation}
H_{\rm wire}(t)= \sum_{n,n'} H_{nn'}(t) c^{\dag}_n c^{\phantom{\dag}}_{n'}\;.
\end{equation}
The fermion operators $c_n$, $c_n^{\dag}$ annihilate and create,
respectively, an electron in the orbital $|n\rangle$.  The influence
of an applied ac-field with frequency $\Omega=2\pi/{\cal T}$
results in a periodic time-dependence of the Hamiltonian:
$H_{nn'}(t+{\cal T})=H_{nn'}(t)$.  The leads are modeled by ideal
electron gases, $H_\mathrm{leads}=\sum_q \epsilon_{q} (c^{\dag}_{Lq}
c^{\phantom{\dag}}_{Lq} + c^{\dag}_{Rq} c^{\phantom{\dag}}_{Rq})$,
where $c_{Lq}^{\dag}$ ($c_{Rq}^{\dag}$) creates an electron in the
state $|Lq \rangle$ ($|Rq \rangle$) in the left (right) lead.  The
tunneling Hamiltonian
\begin{equation}
H_{\rm contacts} = \sum_{q} \left( V_{Lq} c^{\dag}_{Lq} c^{\phantom{\dag}}_1
+ V_{Rq} c^{\dag}_{Rq} c^{\phantom{\dag}}_N
\right) + \mathrm{h.c.}
\end{equation}
establishes the contact between the sites $|1\rangle$, $|N\rangle$
and the respective lead, as sketched in Fig.~\ref{fig:levels}.
Below, we shall assume within a so-termed wide-band limit that the
coupling strengths $\Gamma_{\ell}=2\pi\sum_q |V_{\ell q}|^2
\delta(\epsilon-\epsilon_q)$, $\ell=L,R$, are energy independent.
To specify fully the dynamics, we choose as an initial condition
for the left/right lead a grand-canonical electron ensemble at
temperature $T$ and electro-chemical potential $\mu_{L/R}$,
respectively. An applied voltage $V$ maps to a chemical potential
difference $\mu_R-\mu_L=eV$, where $-e$ is the electron charge.

We shall focus on two central transport quantities: the time-dependent
electrical currents through the two contacts and their fluctuations.
The current operators are given by the negative time variation
of the electron numbers in the leads, multiplied by the electron
charge $-e$, $I_{\ell}(t) = ie [H(t),N_{\ell}]/\hbar$, where
$N_\ell=\sum_q c^{\dag}_{\ell q}c^{\phantom{\dag}}_{\ell q}$ denotes
the electron number in lead $\ell$.

\sect{Generalized Floquet approach}
For the evaluation of correlation functions, we
work in the Heisenberg picture and derive the annihilation operators
at long times by a Floquet ansatz \cite{Grifoni1998a}.
From the Hamiltonian $H(t)$ follow the Heisenberg equations
for the lead operators, with the wire operators appearing in an
inhomogeneity.  In an integrated form they read
\begin{displaymath}
c_{Lq}(t)=c_{Lq}(t_0)e^{-i\epsilon_q(t-t_0)/\hbar}
-\frac{iV_{Lq}}{\hbar}\!\!\int\limits_0^{t-t_0}\!\!\! d\tau\,
e^{-i\epsilon_q\tau/\hbar} c_1(t-\tau)
\end{displaymath}
and $c_{Rq}(t)$ accordingly.  Inserting this into the Heisenberg equations for
the wire operators yields
\begin{align}
\dot c_{1/N} =& -\frac{i}{\hbar} \sum_{n'} H_{1/N,n'}(t)\, c_{n'}
             -\frac{\Gamma_{L/R}}{2\hbar}c_{1/N} + \xi_{L/R}(t), \nonumber \\
\dot c_n =& -\frac{i}{\hbar} \sum_{n'} H_{nn'}(t)\, c_{n'}\,,
\quad n=2,\ldots,N-1.
\label{eq:c}
\end{align}
Owing to the wide-band limit, the dissipative terms are memory free.
Within the chosen grand canonical ensembles the operator-valued
Gaussian noise
$\xi_{\ell}(t)=-(i/\hbar)\sum_q
V^*_{\ell q}e^{-i\epsilon_q(t-t_0)/\hbar}c_{\ell q}(t_0)$
obeys
\begin{align}
\label{eq:xi}
\langle\xi_\ell(t)\rangle =  {}& 0,
\\
\label{eq:xi2}
\langle\xi^\dagger_\ell(t)\,\xi_{\ell'}(t')\rangle
 = {}& \delta_{\ell\ell'}\frac{\Gamma_\ell}{2\pi\hbar^2}
\int\!\! d\epsilon\, e^{i\epsilon(t-t')/\hbar}f_\ell(\epsilon),
\end{align}
where $f_\ell(\epsilon)=(1+\exp[(\epsilon-\mu_\ell)/k_BT])^{-1}$
denotes the Fermi function at temperature $T$ and
chemical potential $\mu_\ell$, $\ell=L,R$.
The current operator then assumes the form
\begin{equation}
\label{eq:I(t)}
I_L(t)=\frac{e}{\hbar}\Gamma_Lc_1^\dagger(t)c_1(t)
       -e\big\{c_1^\dagger(t)\xi_L(t)+\xi_L^\dagger(t)c_1(t)\big\}.
\end{equation}

Before solving the inhomogeneous set of equations \eqref{eq:c},
let us first analyze the corresponding homogeneous equations. They
are linear and possess time-dependent, $\mathcal{T}$-periodic
coefficients. Thus, it is possible to construct a complete
solution with the help of the Floquet ansatz $|\psi(t)\rangle =
\exp[(-i\epsilon_\alpha/\hbar-\gamma_\alpha)\, t]\,
|u_\alpha(t)\rangle$. The Floquet states
$|u_{\alpha}(t)\rangle=\sum_k |u_{\alpha k}\rangle\exp(-ik\Omega
t)$ obey the time-periodicity of the differential equations and
fulfill in a Hilbert space that is extended by a periodic time
coordinate the eigenvalue equation
\begin{equation}
\Big(\mathcal{H}(t) - i\Sigma -i\hbar\frac{d}{dt}\Big)|u_{\alpha}(t)\rangle
=( \epsilon_{\alpha} -  i\hbar\gamma_{\alpha}) |u_{\alpha}(t)\rangle ,
\label{eq:Fs}
\end{equation}
where $\mathcal{H}(t)=\sum_{n,n'} |n \rangle H_{nn'}(t) \langle n'
|$ and $2 \Sigma =|1\rangle\Gamma_L\langle 1 | + |N
\rangle\Gamma_R\langle N |$. Because the eigenvalue
equation \eqref{eq:Fs} is non-Hermitian, its eigenvalues
$\epsilon_{\alpha} - i\hbar\gamma_{\alpha}$ are generally complex
valued and the (right) eigenvectors are not mutually orthogonal.
Therefore, we need to solve also the adjoint Floquet equation
yielding again the same eigenvalues but providing the adjoint
eigenvectors $|u_\alpha^+(t)\rangle$. It can be shown that the
Floquet states $|u_\alpha(t)\rangle$ together with the adjoint
states $|u_\alpha^+(t)\rangle$ form at equal times a complete
bi-orthogonal basis: $\langle u^+_{\alpha}(t)|u_{\beta}(t)\rangle =
\delta_{\alpha\beta}$ and $\sum_{\alpha} |u_{\alpha}(t)\rangle \langle
u^+_{\alpha} (t)|= \mathbf{1}$
\cite{on_orthonormal}\nocite{Jung1990a}.  For
$\Gamma_{L/R}=0$, both $|u_\alpha(t)\rangle$ and
$|u_\alpha^+(t)\rangle$ reduce to the usual Floquet states.

The Floquet states $|u_\alpha(t)\rangle$ allow to write the
general solution of Eq.~\eqref{eq:c} in closed form.  In the
asymptotic limit $t_0\to -\infty$, it reads
\begin{equation}
\label{eq:c(t)}
\begin{split}
c_n(t)=\sum_{\alpha}\int\limits_0^\infty
& d\tau\,
  \langle n|u_\alpha(t)\rangle
  e^{(-i\epsilon_\alpha/\hbar-\gamma_\alpha)\tau}
  \langle u_\alpha^+(t-\tau)| \\
& \times\big\{
  |1\rangle\xi_L(t-\tau)+|N\rangle\xi_R(t-\tau)\big\} .
\end{split}
\end{equation}
To obtain the current $\langle I_L(t)\rangle$, we insert the operator
\eqref{eq:c(t)} into Eq.\ \eqref{eq:I(t)} and use the expectation
values \eqref{eq:xi2}. We then find that the current assumes the
commonly expected ``scattering form'' \cite{Datta1995a} but with
\textit{time-dependent} transmission probabilities and, as well, an additional
contribution that accounts for a $\mathcal{T}$-periodic charging of
the wire.
The latter does not contribute to the time-averaged current
$\bar I=\int_0^\mathcal{T}dt\langle I_{L}(t)\rangle/\mathcal{T}$
so that we obtain as a first result
\begin{equation}
\bar I = \frac{e}{2\pi\hbar}\sum_k\int\! d\epsilon \,
\Big\{ T_{LR}^{(k)}(\epsilon) f_R (\epsilon)
     - T_{RL}^{(k)}(\epsilon) f_L (\epsilon) \Big\} .
\label{eq:I}
\end{equation}
Owing to charge conservation it equals the average current through the
right contact.  The coefficients
\begin{equation}
T_{LR}^{(k)}(\epsilon)=\Gamma_L\Gamma_R
\big|G_{1N}^{(k)}(\epsilon)\big|^2
\end{equation}
can be interpreted as the probability that an electron with energy
$\epsilon$ is transmitted from the right lead to the left lead under
the absorption of $k$ photons, respectively the emission of $-k$
photons when $k<0$.
Note that the sum runs over all integers $k$ 
corresponding to different conduction channels that contribute
independently to the average current $\bar I$.
The retarded Green function
\begin{equation}
\label{eq:G}
G_{nn'}^{(k)}(\epsilon)
=\sum_{\alpha,k'}
 \frac{\langle n|u_{\alpha,k'+k}\rangle\langle u_{\alpha,k'}^+|n'\rangle}
      {\epsilon-(\epsilon_\alpha+k'\hbar\Omega-i\hbar\gamma_\alpha)}
\end{equation}
describes the propagation of an electron from orbital $|n'\rangle$ to
orbital $|n\rangle$.  We emphasize that generally
$|G_{1N}^{(k)}(\epsilon)| \neq |G_{N1}^{(k)}(\epsilon)|$ for a driven
system.
An expression for the time averaged current similar to Eq.\
\eqref{eq:I} has been proposed in Ref.\ \cite{Datta1992a} without
providing an explicit form for the Green function in terms of
generalized Floquet states.

Next, we address the main topic of this work, namely the current noise
given by the auto-correlation function $S_L(t,\tau)= \frac{1}{2}
\langle [\Delta I_L(t),\Delta I_L(t+\tau)]_+\rangle$ of the current
fluctuation operator $\Delta I_L(t) = I_L(t)-\langle I_L(t)\rangle$.
It can be shown that $S_L(t,\tau)=S_L(t+\mathcal{T},\tau)$ shares the
time-periodicity of the driving.  Therefore, it is possible to
characterize the noise level by the time-averaged noise power at zero
frequency, $\bar S_L= \int d\tau \int_0^\mathcal{T}dt \,
S_L(t,\tau)/\mathcal{T}$.  Since the total charge is conserved, the
zero-frequency noise is identical at both the left and the right
contact, i.e.\ $\bar S_L=\bar S_R=\bar S$.
After some extensive algebra we obtain our central result
\begin{widetext}
\begin{align}
\nonumber
\bar S
= \frac{e^2}{2\pi\hbar} \Gamma_L \Gamma_{R} \sum_k  \int\!\! d\epsilon\,
\Big\{ &
  \Gamma_L \Gamma_{R} \Big| \sum_{k'} G_{N1}^{(k'-k)}(\epsilon+k\hbar\Omega)
   G_{N1}^{(k')}(\epsilon)^* \, \Big|^2 f_L(\epsilon) \bar f_L(\epsilon+k\hbar\Omega)
\\
\label{eq:S}
+& \Big| G_{1N}^{(-k)}(\epsilon+k\hbar\Omega) + i \Gamma_L \sum_{k'} 
        G_{1N}^{(k'-k)}(\epsilon+k\hbar\Omega)G_{11}^{(k')}(\epsilon)^* \Big|^2
    f_{L}(\epsilon){\bar f}_{R}(\epsilon+k\hbar\Omega)
\Big\}
\\
\nonumber
&\hspace{-10ex}+\,\text{same terms with the replacement $(L,1)\leftrightarrow(R,N) $} 
, 
\end{align}
\end{widetext}
where we have defined ${\bar f}_{L/R}=1-f_{L/R}$.
The key results \eqref{eq:I} and \eqref{eq:S} contain as special cases
prior findings: In the absence of any driving, the Floquet eigenvalues
$\epsilon_\alpha-i\hbar\gamma_\alpha$ reduce to the complex-valued
eigenenergies; this implies $G_{nn'}^{(k)}=0$ for all $k\neq0$,
yielding the transmission probability for an electron with energy $E$
of $T(E)=\Gamma_L \Gamma_R |G_{N1}^{(0)}(E)|^2$. Thus, the quantities
$\bar I$ and $\bar S$ agree with the well-known expressions obtained
within the time-independent, non-driven scattering approach
\cite{Blanter2000a}.  In the limit of a weak system-lead coupling but
arbitrary driving strength, the average current \eqref{eq:I} coincides
also with the corresponding result of a recent master equation
approach \cite{Lehmann2002b}.

\sect{Current and noise suppression}
As a simple, yet nontrivial application, we consider a wire
composed of $N$ orbitals as sketched in Fig.~\ref{fig:levels}.
Each orbital is coupled to its nearest neighbors by a hopping
matrix element $\Delta$.  The on-site energies are modulated
by the influence of the ac-field, $H_{nn}(t)=E_n-A \cos(\Omega
t)(N+1-2n)/2$, $n=1,\ldots,N$.  The energy $A$ equals the electric
field strength multiplied by the electron charge $-e$ and the
distance between two neighboring sites.  The wire is assumed to
couple equally to both leads, $\Gamma_L=\Gamma_R=\Gamma$,
and the laser frequency is far off resonance, $\Omega
= 5 \Delta/\hbar$.
For a molecular wire, a typical value for the hopping matrix element
$\Delta$ and the coupling strength $\Gamma$ is $0.1\,$eV leading to a
current unit $e\Gamma/\hbar \simeq 25\,\mu\mathrm{A}$, while the laser
frequency lies in the optical regime. For a distance of $2\,${\AA}
between two neighboring sites, a driving amplitude $A=\Delta$
corresponds to an electric field strength of roughly
$5\times10^6\,\mathrm{V/cm}$.
For the evaluation of the current $\bar I$ and the noise $\bar S$,
we restrict ourselves to zero temperature.  Then, the
Fermi functions turn into step functions and the energy integrals
in Eqs.~\eqref{eq:I} and \eqref{eq:S} can be carried out analytically.
This limit is physically well justified for molecular wires at room
temperature and for quantum dots at helium temperature since in
both cases, thermal electron excitations do not play a significant
role.

\begin{figure}[t]
\includegraphics[width=0.95\columnwidth]{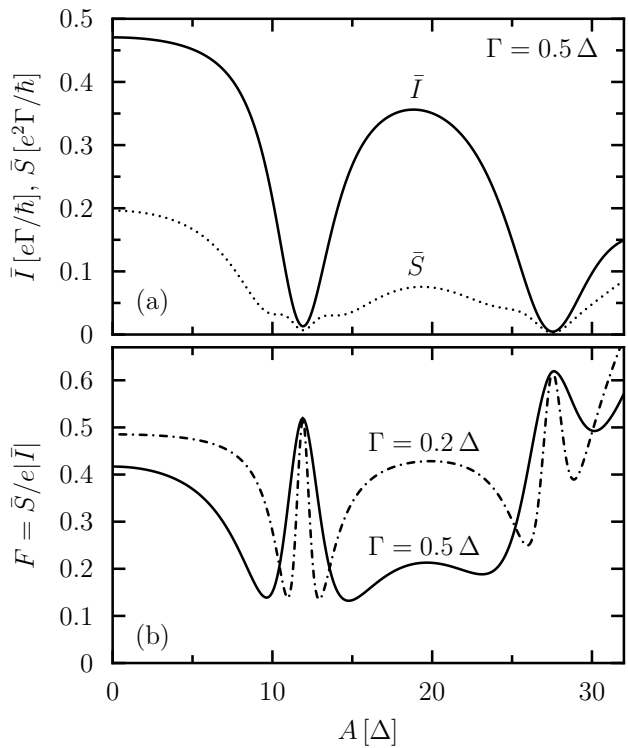}
\caption{\label{fig:IS}
Time-averaged current $\bar I$ and zero-frequency noise $\bar S$
(a) as a function of the driving amplitude $A$ for a wire
with $N=3$ sites with on-site energies $E_n=0$ and chemical potentials
$\mu_R=-\mu_L=25\Delta$.  The other parameters read $\Omega = 5
\Delta/\hbar$ and $\Gamma = 0.5\Delta$.
Panel (b) displays the Fano $F$ factor for these parameters (full line)
and for smaller wire-lead coupling (dash-dotted line).}
\end{figure}%
Figure \ref{fig:IS}a depicts the dc-current and the zero-fre\-quen\-cy
noise for a wire with $N=3$ sites with equal on-site energies and a
relatively large applied voltage.  As a remarkable feature, we find
that for certain values of the field amplitude $A$, the current drops
to a value of some percent of the current in the absence of the field
\cite{Lehmann2002c}.
A perturbation theory for the Floquet equation \eqref{eq:Fs} for
$\Delta,\Gamma\ll\hbar\Omega$ yields that the driving results in a
renormalized hopping matrix element $\Delta\to\Delta_\mathrm{eff}=
\Delta J_0(A/\hbar\Omega)$, where $J_0$ denotes the zeroth-order
Bessel function.  Then, $G_{1N}$ and $G_{N1}$ vanish
if the condition $J_0(A/\hbar\Omega)=0$ is fulfilled
\cite{Grossmann1992a}.  Consequently, the dc-current \eqref{eq:I}
and the zero-frequency noise \eqref{eq:S} also vanish.

The \textit{relative} noise strength can be characterized by the
so-called Fano factor $F=\bar S/e|\bar I|$ depicted in Fig.~\ref{fig:IS}b.
Interestingly enough, we find that the Fano factor exhibits as a function of
the driving amplitude $A$ both a sharp maximum at current suppression and two
pronounced minima nearby.  For a sufficiently large voltage, the Fano factor
assumes at the maximum the value $F\approx 1/2$.
Once the driving amplitude is of the order of the applied
voltage, however, the Fano factor becomes much larger.
The relative noise minima are
distinct and provide a typical Fano factor of $F\approx0.15$.
Reducing the coupling to the leads renders these phenomena even more
pronounced since then the suppressions occur in a smaller interval of
the driving amplitude, cf.\ Fig.\ \ref{fig:IS}b. The overall behavior
is robust in the sense that approximately the same values for the
minima and the maximum are also found for larger wires, different
driving frequencies, different coupling strengths, and slightly
modified on-site energies, provided that $\Delta,\Gamma,E_n \ll
\hbar\Omega$ and that the applied voltage is sufficiently large.
The qualitative behavior can again be understood within a perturbative
approach.  With increasing driving amplitude, a crossover from
$\Delta_\mathrm{eff} \gg \Gamma$ to $\Delta_\mathrm{eff} \ll \Gamma$
at the current suppression occurs.  Both limits correspond to the
transport through a symmetric double barrier and, therefore, are
characterized by $F\approx1/2$ \cite{Blanter2000a}.  At the crossover
$\Delta_\mathrm{eff} \approx \Gamma$ the effective barriers vanish
and, consequently, the Fano factor assumes its minimum.

In summary, we have put forward with Eqs.\ \eqref{eq:I} and
\eqref{eq:S} new and appealing expressions for the time-averaged
current and the zero-frequency noise for the electron transport
through ac-driven nanoscale systems.
A main finding is that in molecular wires and coherently coupled
quantum dots, the influence of an ac-field can be used to selectively
suppress both the current and its noise.  Investigating the
\textit{relative} noise level characterized by the Fano factor has
revealed that the current suppression is accompanied by a noise
maximum and two remarkably low minima.  These phenomena can be used to
devise novel current sources with \textit{a priori} controllable noise
levels.

%
We thank G.-L.\ Ingold for helpful discussions.
This work has been supported by the Volkswagen-Stiftung under
Grant No.\ I/77 217, a Marie Curie fellowship of the European
community program IHP under contract No.\ HPMF-CT-2001-01416
(S.C.), and the DFG Sonderforschungsbereich 486.


\end{document}